\documentclass[twocolumn,aps,prl,superscriptaddress,nofootinbib]{revtex4-1}
\usepackage{graphicx}
\usepackage{color}
\graphicspath{{figures/}{fig/}}
\usepackage{amsmath}
\usepackage{bm}
\usepackage{slashed}
\usepackage{epsfig}
\usepackage{amsfonts}
\usepackage{epstopdf}
\usepackage{hyperref}
\usepackage{bbm}
\usepackage{textcomp}

\newcommand{\sect}[1]{{\it \textbf{#1} ---}}
\newcommand{\ui}{\mathrm{i}}
\newcommand{\ud}{\mathrm{d}}
\begin{document}
\title{A Systematic and Efficient Method to Compute Multi-loop Master Integrals}

\author{Xiao Liu}
\email{xiao6@pku.edu.cn}
\affiliation{School of Physics and State Key Laboratory of Nuclear Physics and
Technology, Peking University, Beijing 100871, China}
\author{Yan-Qing Ma}
\email{yqma@pku.edu.cn}
\affiliation{School of Physics and State Key Laboratory of Nuclear Physics and
Technology, Peking University, Beijing 100871, China}
\affiliation{Center for High Energy physics, Peking University, Beijing 100871, China}
\affiliation{Collaborative Innovation Center of Quantum Matter,
Beijing 100871, China}
\author{Chen-Yu Wang}
\email{wangcy@pku.edu.cn}
\affiliation{School of Physics and State Key Laboratory of Nuclear Physics and
Technology, Peking University, Beijing 100871, China}

\date{\today}

\begin{abstract}
We propose a novel method to compute multi-loop master integrals by constructing and numerically solving a system of ordinary differential equations, with almost trivial boundary conditions. Thus it can be systematically applied to problems with arbitrary kinematic configurations. Numerical tests show that our method can not only achieve results with high precision, but also be much faster than the only existing systematic method sector decomposition. As a by product, we find a new strategy to compute scalar one-loop integrals without reducing them to master integrals.
\end{abstract}

\maketitle
\allowdisplaybreaks

\sect{Introduction}
With the continuous improvement of statistics and experimental systematics at the Large Hadron Collider, the aim of  testing the particle physics Standard Model and discovering new physics strongly demands theoretical predictions to also improve uncertainty to the same level. For many important processes, high order perturbative calculations are needed to this end.
At the one-loop order, thanks to the improvement of traditional tensor reduction \cite{Denner:2005nn} and the  development of unitarity-based reduction  \cite{Britto:2004nc,Ossola:2006us,Giele:2008ve}, one can  efficiently express scattering amplitudes in terms of linear combinations of master integrals (MIs). As the computation of one-loop MIs is a solved problem \cite{tHooft:1978jhc,Passarino:1978jh,vanOldenborgh:1989wn},  one-loop calculations can now be done automatically.  Expressing multi-loop scattering amplitudes in terms of MIs is also possible using such as the integration-by-parts (IBP) reduction~\cite{Chetyrkin:1981qh,Laporta:2001dd,Studerus:2009ye,Lee:2012cn,Smirnov:2014hma,vonManteuffel:2014ixa} or the unitarity-based multi-loop reduction \cite{Gluza:2010ws,Kosower:2011ty,Mastrolia:2012wf,Badger:2012dp,Mastrolia:2013kca,Larsen:2015ped,Ita:2015tya,Badger:2016ozq,Mastrolia:2016dhn,Georgoudis:2016wff,Peraro:2016wsq,Abreu:2017idw,Abreu:2017xsl}. Then, one of the main obstacles for multi-loop calculation is the computation of multi-loop MIs.

We take two recent studies in literature as examples to demonstrate how hard the multi-loop MIs computation is. One example is a  two-loop calculation of pseudoscalar quarkonium inclusive decay \cite{Feng:2017hlu}, where the computational expense of MIs is about ${\cal O}(10^5)$ CPU core-hour. Another example is a calculation of four-loop nonplanar cusp anomalous dimension \cite{Boels:2017skl}. The reduction of  amplitudes to MIs in this problem has been done much earlier in Ref. \cite{Boels:2015yna}, yet the computation of these MIs is very challenging. The final numerical result obtained in Ref.  \cite{Boels:2017skl} has uncertainty as large as 10\% , which we believe is already the best precision that one can get with a tolerable computational expense.

Currently, the only method that can systematically compute any MI is the sector decomposition~\cite{Binoth:2000ps}. Unfortunately, this method is extremely time-consuming, besides that it is hard to achieve high precision. Mellin-Barnes representation ~\cite{Smirnov:1999gc} is another widely used method, yet it has difficulty to deal with non-planar diagrams, at least not in a systematic way (see Ref. \cite{Dubovyk:2016aqv} and references therein for recent progress). The differential equation (DE) method~\cite{Kotikov:1990kg,Bern:1992em,Remiddi:1997ny,Gehrmann:1999as} is a powerful tool to compute multi-loop MIs, which bases on the fact that derivation of a MI with respect to its kinematic variables  (including Mandelstam variables and internal masses) can be re-expressed as a linear combination of MIs using aforementioned reductions.
For simple problems, DE method can give analytical results thanks to the introduction of canonical form  \cite{Henn:2013pwa, Lee:2014ioa,Adams:2017tga}; while for complicated problems, one can solve DEs numerically to achieve results with high precision (see \cite{Caffo:2008aw,Czakon:2008zk,Mueller:2015lrx,Lee:2017qql} and references therein).
However, it needs input of boundary conditions (BCs) of MIs evaluated at another set of kinematic configurations. As there is no general rule to obtain BCs for arbitrary problems at present, one needs to find good BCs case by case, which makes it hard for DE method to be systematical.
In practice, sector decomposition method is employed in Ref. \cite{Feng:2017hlu}, and both sector decomposition method and Mellin-Barnes representation method are employed in Ref. \cite{Boels:2017skl}.

In this Letter, we develop a novel method to compute multi-loop MIs by constructing and solving a system of ordinary differential equations (ODEs). Advantages of our method are as follows: 1) Our BCs are fully massive vacuum integrals with a single mass scale, which are much simpler to compute and have been well studied in literature \cite{Luthe:2015ngq}. As a result, our method can be systematically applied to any complicated problem; 2) ODEs can be numerically solved efficiently to high precision, no matter how many mass scales are involved in the problem; 3) Computing MIs with complex kinematic variables is very easy in our method, while it could be hard for other methods (note that introducing imaginary part to kinematic variables is needed for many purposes, e.g., to describe particle decay or to study the S-matrix theory). Numerical tests show that our method can be much faster than the only existing systematic method sector decomposition. As a by product, we find a new strategy to compute scalar one-loop integrals in arbitrary spacetime dimensions without reducing them to MIs.

\sect{The method}
Let us introduce a dimensionally regularized $L$-loop MI,
\begin{align}\label{eq:int}
I(D;\{\nu_\alpha\};\eta)\equiv\int\prod_{i=1}^{L}\frac{\ud^D\ell_i}{\ui\pi^{D/2}}\prod_{\alpha=1}^{N}\frac{1}{(\mathcal{D}_\alpha+\ui\eta)^{\nu_{\alpha}}}\, ,
\end{align}
where $D$ is the spacetime dimension, $\mathcal{D}_\alpha\equiv q_\alpha^2-m_\alpha^2$ are usual Feynman propagators, and $q_\alpha$ are linear combinations of loop momenta $\ell_i$ and external momenta $p_i$. The actual integral that we want to get is
\begin{align}
I(D;\{\nu_\alpha\};0)\equiv \lim_{\eta\to0^+} I(D;\{\nu_\alpha\};\eta),
\end{align}
with $0^+$ defining the causality of Feynman amplitudes. In the following, we will suppress the dependence on $D$ and $\{\nu_\alpha\}$ whenever it does not introduce any confusion.

We set up ODEs by differentiating MIs with respect to $\eta$ and then re-expressing them in terms of MIs, which results in
\begin{align}\label{eq:DE}
\frac{\partial}{\partial\eta}\vec{I}(\eta)=A(\eta)\vec{I}(\eta)\, ,
\end{align}
where $\vec{I}(\eta)$ is the vector of a complete set of $m$ MIs and $A(\eta)$ is the $m\times m$ coefficient matrix. To obtain MIs at $\eta=0^+$, we solve the ODEs with BCs chosen at $\eta=\infty$. As we will show, BCs are simply vacuum integrals with equal masses, which can be computed rather easily. Considering also that numerically solving these ODEs is well-studied mathematical problem,
our method provide a systematic and efficient way to compute multi-loop MIs to high precision.

\sect{Boundary conditions}
Before studying BCs rigorously, let us explain the idea of choosing BCs at $\eta=\infty$. With a sufficiently large imaginary part in all denominators, we expect all kinematic variables to be negligible because they are finite. Thus we should be able to set both internal masses $m_\alpha$ and external momenta $p_i$ to zero at the boundary, which results in simple vacuum integrals with equal masses. The only loophole in this argument is that, as loop momenta $\ell_i$ can be arbitrarily large, it is not obvious that $\ell_i\cdot p_j$ are negligible comparing with $\eta$ even if $\eta\to\infty$. The loophole can be fixed by studying its Feynman parametric representation, and then our na\"{\i}ve expectation holds in general.

We assume $\nu_{\alpha}>0$ for all $\alpha$ in Eq.~\eqref{eq:int} to simplify our discussion, although our final conclusion is unchanged even without this condition. Then the Feynman parametric representation of Eq.~\eqref{eq:int} is given by
\begin{align}\label{eq:feynman}
I(\eta)=&\,(-1)^{\nu}\frac{\Gamma \left(\nu-LD/2\right)}{\prod_i\Gamma(\nu_{i})}\int\prod_\alpha (x_\alpha^{\nu_{\alpha}-1}\ud x_\alpha)\nonumber\\
&\times\delta\bigg( 1-\sum_j x_j\bigg)\frac{\mathcal{U}^{-D/2}}{(\mathcal{F/U}-\ui\eta)^{\nu-LD/2}}\, ,
\end{align}
where $\mathcal{U}$ and $\mathcal{F}$ are so-called graph polynomials that can be related to the spanning 1-tree and 2-tree of the original Feynman diagram, respectively (see e.g. Ref.~\cite{Bogner:2010kv}), and $\nu$ is short for $\sum_\alpha\nu_{\alpha}$. All kinematic variables are incorporated in $\mathcal{F}$, leaving $\mathcal{U}$ depending only on Feynman parameters.

An important observation is that $|\mathcal{F/U}|$ is bounded in the open interval of Feynman parameter space. To show this, we express $\mathcal{F}=\sum_i \mathcal{F}_i$ and $\mathcal{U}=\sum_i \mathcal{U}_i$, where $\mathcal{F}_i$ and $\mathcal{U}_i$ are monomials in Feynman parameters. By definition, a 2-tree can be generated by a 1-tree, i.e. there exists a pair of indexes $j$ and $k$ so that $\mathcal{F}_i=t_i\mathcal{U}_j x_k$, where $t_i$ is the kinematic part of $\mathcal{F}_i$. We then have $|\mathcal{F}_i|<|t_i||\mathcal{U}_i|<|t_i||\mathcal{U}|$ and $|\mathcal{F}|<\sum_i|t_i||\mathcal{U}|$, where we have used the fact that $\mathcal{U}_i$ are positive definite in the open interval. As $\sum_i|t_i|$ is finite, we conclude that $|\mathcal{F/U}|$ is bounded.

Because $|\mathcal{F/U}|$ is bounded, $\mathcal{F/U}$ in the denominator of Eq.~\eqref{eq:feynman} can be neglected as $\eta \to \infty$. This effectively sets all kinematic variables to zero in the original integral, because $\mathcal{F}$ includes all kinematic variables. The result is a fully massive vacuum integral $I^{\mathrm{bub}}(\eta)$ which shares the same internal topology as the original integral. Because this is a single scale integral, the $\eta$ dependence can be factorized out, which results in a relation
\begin{align}\label{eq:relation}
I(\eta)=\eta^{LD/2-\nu}\Big[I^{\mathrm{bub}}(1)+\mathcal{O}(\eta^{-1})\Big],
\end{align}
where $ I^{\mathrm{bub}}(1)$ can be interpreted as a vacuum integral with equal internal squared masses $m^2=-\ui$.
It is worth mentioning that the object $J(\eta)\equiv\eta^{\nu-LD/2}I(\eta)$ is analytic near $\eta=\infty$ based on the above discussion.

\begin{figure}[htb]
\vspace{-0.5cm}
\begin{center}
\includegraphics[width=0.85\linewidth]{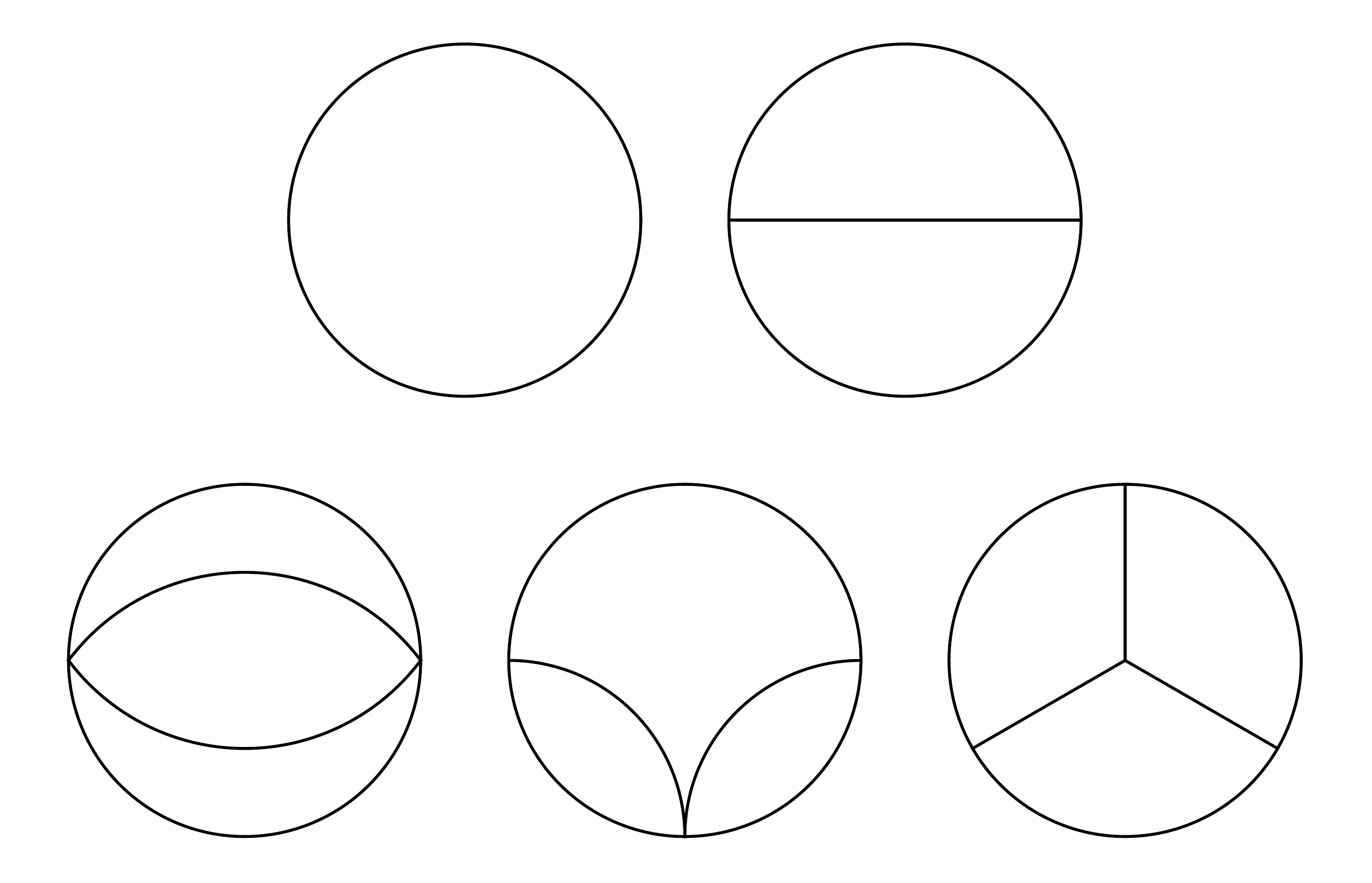}
\caption{\label{fig:bc}
Diagrams of nonfactorizable vacuum master integrals up to 3 loops.}
\end{center}
\vspace{-0.5cm}
\end{figure}

To compute $ I^{\mathrm{bub}}(1)$, we again reduce it to linear combination of corresponding vacuum MIs, diagrams of which up to 3 loops are shown in Fig.~\ref{fig:bc}. Computation of these vacuum MIs is well studied, with analytical results available up to 3 loops~\cite{Davydychev:1992mt,Broadhurst:1998rz,Kniehl:2017ikj} (see \cite{threeloops} and references therein for some pioneering works) and numerical results available up to 5 loops~\cite{Schroder:2005va,Luthe:2015ngq,Luthe:2017ttc}. We therefore conclude that the computation of BCs in our method is a solved problem.

\sect{Solving ODEs numerically}
Knowing BCs, solving the ODEs numerically to obtain MIs at $\eta=0^+$ is a well-studied mathematical problem. The solution can be obtained efficiently to high precision.

\begin{figure}[htb]
\begin{center}
\includegraphics[width=0.85\linewidth]{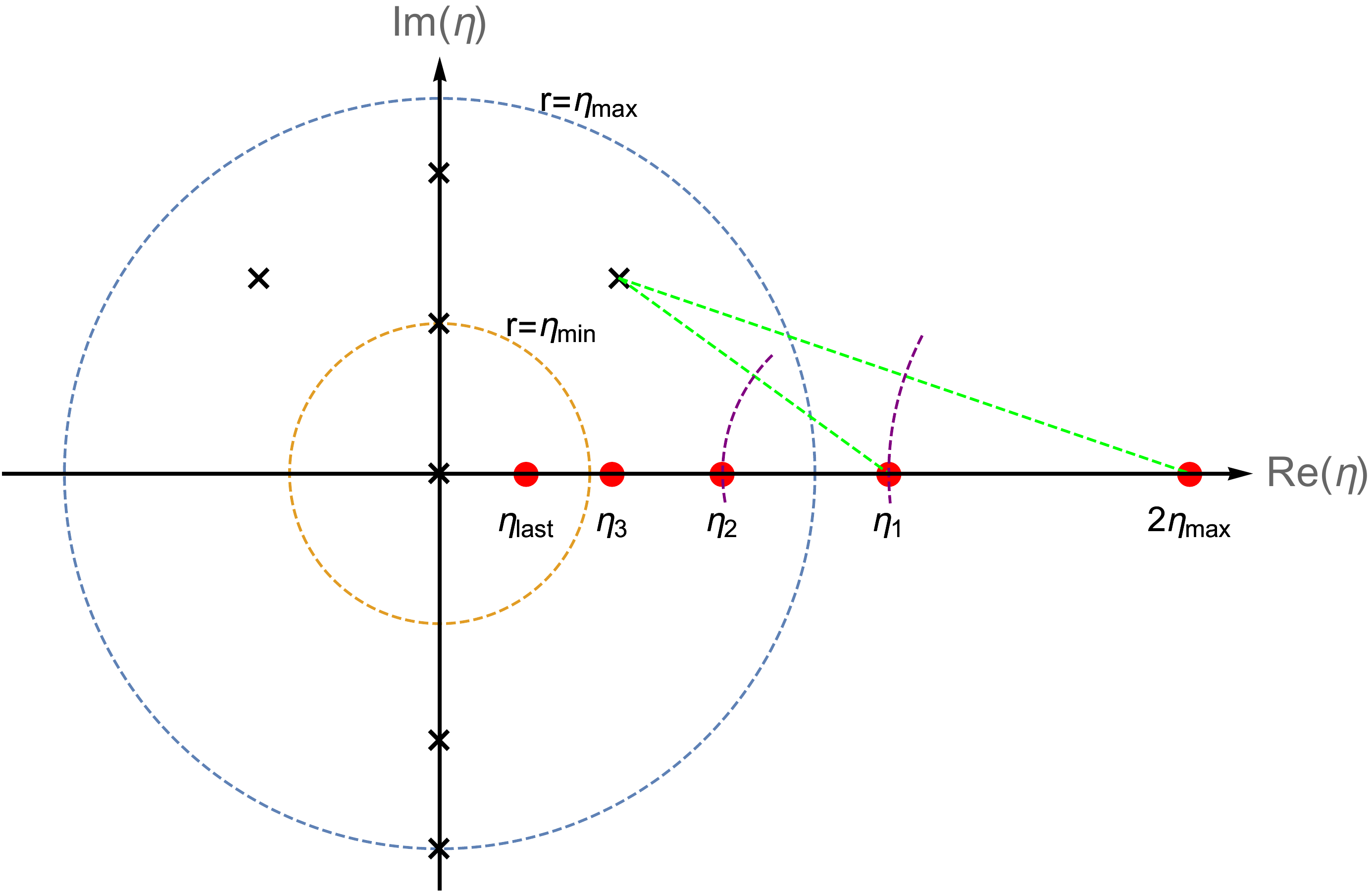}
\caption{\label{fig:pole}
An example distribution of singularities of ODEs. Singularities are labeled as crosses. Solid dots are points where to expand the ODEs.}
\end{center}
\vspace{-0.5cm}
\end{figure}

Singularities, which restrict the convergence domain of Taylor expansion or asymptotic expansion, play essential roles in the process of solving ODEs. For cases with only real kinematic variables, most singularities of our ODEs are located on the imaginary axis of the $\eta$ complex plane. These singularities are usually branch points of some MIs. Besides, ODEs have additional singular points not on the imaginary axis, which are not singularities of MIs. This can be easily understood from the definition of MIs in Eq.~\eqref{eq:int}, where no propagator could vanish if $\eta$ had a finite real part.  An example distribution of singularities of ODEs is shown in Fig~\ref{fig:pole}, where there are branch cuts on the imaginary axis. For later convenience, we define $\eta_{\mathrm{max}}=\mathrm{max}\{|s_i|\}$ and $\eta_{\mathrm{min}}=\mathrm{min}\{|s_i|\}$, where $s_i$ runs over all singularities except $0$ and $\infty$.

Having located all singular points, an algorithm to solve the system of ODEs can be roughly divided into three steps:
\begin{itemize}
\item
Step 1: Using ODEs to take Taylor expansion of modified MIs around $\eta=\infty$ and to obtain values of MIs at a point outside of the large circle, which we choose $\eta=\eta_0=2\eta_\mathrm{max}$ as shown in Fig.~\ref{fig:pole}.
\item
Step 2: Using ODEs to take Taylor expansion of MIs around $\eta=\eta_j \, (j=0, 1, \cdots)$ and obtain their values  at $\eta=\eta_{j+1}$. Do it repeatedly until $\eta_{j+1}$ is inside of the small circle, denoting $\eta=\eta_\mathrm{last}$ in Fig.~\ref{fig:pole}.
\item
Step 3: Using ODEs to take asymptotic expansion of MIs around $\eta=0$ and matching with their values at $\eta=\eta_\mathrm{last}$ to obtain dimensionally regularized MIs at $\eta=0^+$.
\end{itemize}

Modified MIs at the step 1 are combinations $J_k(\eta)=\eta^{\nu_k-LD_k/2} I_k(\eta)$, which are analytic around $\eta=\infty$ as pointed out above. From BCs at $\eta=\infty$, ODEs tell us their first-order derivatives, based on which ODEs then tell us their second-order derivatives, and so on. Eventually we get the Taylor series at $\eta=\infty$. Taylor expansion at the step 2 can be obtained similarly.

Having obtained the values of $\vec{I}(\eta_\mathrm{last})$ with $\eta_\mathrm{last}<\eta_\mathrm{min}$ from step 2, we then do an asymptotic expansion of MIs at $\eta=0$. We rewrite Eq.~\eqref{eq:DE} as
\begin{align}\label{eq:zero}
\eta\frac{\partial}{\partial\eta}\vec{I}(\eta)=\tilde{A}(\eta)\vec{I}(\eta)\, .
\end{align}
In general, the matrix $\tilde{A}$ can be divergent as $\eta^{-p}$ with positive integer $p$, which is called the Poincar\'{e} rank at $\eta=0$. However, because MIs have integral representation, they can only have regular singularities on the whole complex plane of $\eta$. This enables us to reduce the Poincar\'{e} rank at $\eta=0$ by a transformation (see Ref.~\cite{Lee:2014ioa} and references therein for different algorithms). So, without loss of generality, we can assume that $\tilde{A}(0)$ is finite.
Then the general solution of  Eq.~\eqref{eq:zero} near $\eta=0$ has the form\footnote{Seriously speaking, this is true if and only if that the difference of any two distinguished eigenvalues of $\tilde{A}_0$ is non-integer. See, e.g., Ref~\cite{Wason:1987aa} for dealing with these exceptional cases.}
\begin{align}\label{eq:exp0}
\vec{I}(\eta)=&\, P(\eta)\mathrm{exp}\left\{\tilde{A}(0){\ln}(\eta)\right\}\vec{v}_0\nonumber\\
\equiv&\,\sum_{n=0}^{\infty}P_n \eta^{n+\tilde{A}(0)}\vec{v}_0\, ,\quad \text{for }|\eta|<\eta_\mathrm{min}\, ,
\end{align}
where the matrix $P(\eta)$ has been expanded as $\sum_{n=0}^{\infty}P_n \eta^{n}$ with $P_0$ being the identity matrix. In the above expansion, non-analytical behaviors of $\eta$, like $\eta^{a+b D}$ or $\ln( \eta)$, are contained in the matrix exponential $\mathrm{exp}\left\{\tilde{A}(0){\ln}(\eta)\right\}=\eta^{A(0)}$. Substitute Eq.~\eqref{eq:exp0} into Eq.~\eqref{eq:zero}, we get a system of equations,
\begin{align}
n P_n+[P_n,\tilde{A}_0]=\sum_{k=0}^{n-1}\tilde{A}_{n-k}P_k\, ,
\end{align}
which can be solved recursively to obtain $P_n$ ($n\ge1$). It is clear that $P(\eta)$ is determined completely by the matrix $\tilde{A}(\eta)$, or equivalently by the ODEs. As a consequence, all boundary information are contained in the vector $\vec{v}_0$, which can be determined uniquely by matching the value of $\vec{I}(\eta_\mathrm{last})$ obtained from step 2,
\begin{align}
\vec{I}(\eta_\mathrm{last})=P(\eta_\mathrm{last})\eta_\mathrm{last}^{\tilde{A}(0)}\vec{v}_0\, .
\end{align}
Then we have
\begin{align}\label{eq:solve}
\vec{I}(0)=&\lim_{\eta\rightarrow0^+}\eta^{\tilde{A}(0)}\vec{v}_0\nonumber\\
=&\lim_{\eta\rightarrow0^+}\eta^{\tilde{A}(0)}\left[P(\eta_\mathrm{last})\eta_\mathrm{last}^{\tilde{A}(0)}\right]^{-1}\vec{I}(\eta_\mathrm{last})\, .
\end{align}
For dimensional regularized Feynman integrals, all terms that are non-analytic in $\eta$ vanish in the limit $\eta\rightarrow0^+$, which results in well-defined $\vec{I}(0)$. All infrared divergences emerge in this limit. As a result, our method works regardless of whether propagators in the Feynman integral are massive or massless. In practice, instead of using Eq.~\eqref{eq:solve} directly, a more efficient way is to solve nonhomogeneous ODEs sequentially.

\sect{An one-loop example}
A fascinating feature of our method is that ODEs at one-loop level can be easily constructed for all scalar integrals, not restricted to MIs. We will only consider scalar integrals with nonvanishing Gram determinant, because scalar integrals with vanishing Gram determinant can be easily reduced to the former case.  Using a raising dimensional recurrence relation and a lowering dimensional recurrence relation [see e.g., Eqs.(14) and (18) in Ref. ~\cite{Duplancic:2003tv}], we get
\begin{align}\label{eq:de1}
\frac{\partial}{\partial\eta} I&(D;\{\nu_\beta\};\eta)=\frac{1}{2\eta-\ui C}\Big[{(D-1-\nu)}I(D;\{\nu_\beta\};\eta)\nonumber\\
&+\sum_{\alpha=1}^{N}{ z_\alpha}I(D-2;\{\nu_\beta-\delta_{\alpha\beta}\};\eta)\Big]\, ,
\end{align}
where $C=-\frac{\text{det}(R)}{\text{det}(S)}$ with $\text{det}(S)$ and $\text{det}(R)$ are Gram and modified Cayley determinants of $I(D;\{\nu_\beta\};0)$, respectively, and $z_{\alpha}$ with $\sum_{\alpha=1}^{N}{ z_\alpha}=1$ are solutions of the linear equations $\sum_{\alpha=1}^N R_{\beta\alpha} z_\alpha=C (\beta=1,\cdots, N)$.  It is worth mentioning that both $C$ and $z_{\alpha}$ are independent of $\eta$. Therefore, Eq.~\eqref{eq:de1} has only a purely imaginary singularity at $\eta=\frac{\ui}{2}C$, which is the leading Landau singularity of $I(D;\{\nu_\beta\};\eta)$.

The DE~\eqref{eq:de1} naturally connects a $D$ dimensional scalar integral to some $(D-2)$ dimensional scalar integrals with one less propagator. Thus, we should better choose this set of scalar integrals as basis to solve the system of ODEs. Based on this, we can compute any one-loop scalar integral.

\begin{figure}[htb]
\begin{center}
\begin{minipage}[c]{0.45\textwidth}
\centering
\includegraphics[width=0.8\linewidth]{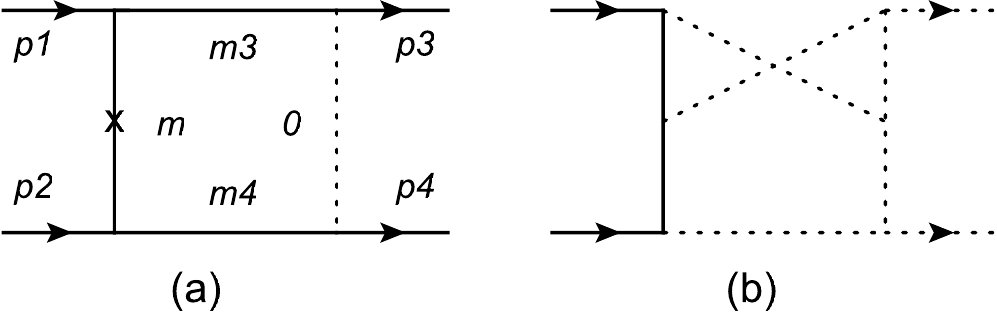}
\caption{\label{fig:1loop}
(a) A 1-loop box diagram; (b) A 2-loop non-planar box diagram.}
\end{minipage}
\end{center}
\vspace{-0.5cm}
\end{figure}

As an example, we compute an infrared-divergent multi-scale  scalar integral shown in Fig.~\ref{fig:1loop}(a), with the cross symbol indicating that the power of the corresponding propagator is two. We set $D=4-2\epsilon$ and express the final results as Laurent series up to ${\cal O}(\epsilon)$.  Kinematic variables are chosen as $p_1^2=1.2,p_2^2=3.1,p_3^2=m_3^2=2.7,p_4^2=m_4^2=7.5,m^2=5.4,t=(p_1-p_3)^2=-1.0$, and $s=(p_1+p_2)^2=(m_3+m_4)^2(1+\delta)=19.2(1+\delta)$. When $|\delta|\ll1$, $s$ is close to the threshold $(m_3+m_4)^2$, and then we have $\eta_\mathrm{min}\approx4.5|\delta|$ and $\eta_\mathrm{max}\approx10.2$. We can run around $3+\log_2(\eta_\mathrm{max}/\eta_\mathrm{min})\approx4.2+1.4\ln(1/|\delta|)$ steps to go from $\eta=\infty$ to $\eta=0^+$. Thus the computational complexity increases slowly as $s$ close to threshold.

By taking Taylor expansion or asymptotic expansion to 30 orders at each step of the running process,  we find that the obtained final numerical results for the above scalar integral have at least 10 correct significant digits for any choice of $\delta$ in the range $10^{-7}\le|\delta|\le1$ . To compare with, we find that the sector decomposition approach \verb"FIESTA4"~\cite{Smirnov:2015mct} can get result with tolerable uncertainty only for $|\delta|\ge10^{-3}$.

\sect{A two-loop example}
We take the non-planar scalar integral shown in Fig~\ref{fig:1loop}(b) as a two-loop example, where solid lines have masses $m^2=1$, dashed lines are massless, $s=4$, and $t=-1$.
This is the most time-consuming MI in the study of Ref.~\cite{Feng:2017hlu}.

For multi-loop integrals, similar relations as Eq.~\eqref{eq:de1} are not available yet. One needs to rely on reduction methods. We use \verb|FIRE5|~\cite{Smirnov:2014hma} to construct ODEs. The resulted ODEs contain 168 MIs. There are totally 26 distinguished singularities of the ODEs, 6 among which are not on the imaginary axis with approximate values $\pm0.629-0.222 \ui,\pm0.359-1.059 \ui,$ and $\pm0.127-0.974 \ui$. With $\eta_\mathrm{min}\approx0.118$ and $\eta_\mathrm{max}=4$, we run 14 steps to go from $\eta=\infty$ to $\eta=0^+$. By taking Taylor expansion or asymptotic expansion to 30 orders at each step, we get numerical result for the non-planar integral in Fig~\ref{fig:1loop}(b),
\begin{align}
&I_\mathrm{np}(4-2\epsilon)=0.0520833\epsilon^{-4}-(0.131616-0.147262\ui)\epsilon^{-3}\nonumber\\
&\,-(0.741857+0.185602\ui)\epsilon^{-2}+(3.73984-4.15756\ui)\epsilon^{-1}\nonumber\\
&\,-(4.75677-12.0749\ui)+(23.9674-55.4214\ui)\epsilon+\cdots\,,
\end{align}
together with results of other 167 integrals in the same sector, all of which have at least 6 correct significant digits. We obtain these numerical results using our \verb"Mathematica" code in a few minutes on a laptop. The efficiency can be further improved  in a future \verb"Fortran" code.
To compute MIs in the same sector using \verb"FIESTA4" in Ref.~\cite{Feng:2017hlu},  it cost about ${\cal O}(10^4)$ CPU core-hour to achieve results with 1\textperthousand~uncertainty \footnote{We thank Yu Jia for letting us know this.}.  Therefore, our method is faster by at least $10^5$ times than \verb"FIESTA4" for this problem. Using finite integrals as bases\cite{vonManteuffel:2017myy} or using other sector decomposition codes, e.g. \verb"pySecDec" \cite{Borowka:2017idc}, may speed up the calculation for this problem, but the efficiency should be still hard to compete with our method.

\sect{Summary and outlook}
In this Letter, we presented a novel method to compute multi-loop MIs, which works by constructing and solving ordinary differential equations of MIs with respect to Feynman prescription parameter $\eta$. On the one hand, the method retains all advantages of usual DE approach, such as efficient for numerical evaluation, stable around threshold region, and flexible to introduce imaginary kinematic variables; and on the other hand, the method overcomes the main difficulty of usual DE approach, with almost trivial boundary conditions. Therefore, our method can compute MIs systematically and efficiently to high precision, which makes it possible to define MIs as new {\it special functions} instead of expressing them in terms of other special functions.

Based on a two-loop example, it was found that our method can be much faster than the only existing systematic method, sector decomposition. With this significant improvement of efficiency, one can now use reasonable computer source to perform phase space integration in multi-loop calculations. To this purpose, one needs to construct and solve ODEs of MIs at each given phase space point, which is very natural in our method.

One thing which we want to clarify is that, to estimate the time consumed in our method, we only counted the time to solve ODEs, but ignored the time to construct ODEs. The reason is as follows. Whenever one wants values of MIs for a multi-loop calculation, one must have expressed scattering amplitudes as linear combination of them by reduction methods. The construction of ODEs of MIs is usually much faster than the full reduction of scattering amplitudes, and thus its time can be neglected. We further note that a new reduction method is proposed in Ref. \cite{Liu:2018dmc}, which may significantly reduce the time to construct ODEs.

Many variants of our method can be constructed easily. The basic idea of our method is to introduce an infinite large value in propagators at the boundary of ODEs, so that all kinematic variables can be dropped out. This aim can be equally met, e.g., by only differentiating the parameter $\eta$ of some propagators in the construction of ODEs. Then to compute corresponding BCs, one again constructs ODEs but with respect to $\eta$ of other propagators.

\sect{Acknowledgments}
We thank Kuang-Ta Chao, Feng Feng, Yu Jia, Zhao Li, Xiaohui Liu, Ce Meng and Yang Zhang for helpful discussions. The work is supported
in part by the National Natural Science Foundation of
China (Grants No. 11475005).



\end{document}